\documentclass[twocolumn,showpacs,prl]{revtex4}

\usepackage{graphicx}

\usepackage{amsmath}

\usepackage{amssymb}

\usepackage{natbib}

\begin{document}

\title{Rheology and Contact Lifetime Distribution in Dense Granular Flows}

\author{Leonardo E.~Silbert${^1}$}

\author{Gary S. Grest$^{2}$}

\author{Robert Brewster$^{3}$}

\author{Alex J. Levine$^{3}$}

\affiliation{$^{1}$Department of Physics, Southern Illinois
  University, Carbondale, IL 62901, USA}

\affiliation{$^{2}$Sandia National Laboratories, Albuquerque, NM
  87185, USA}

\affiliation{$^{3}$Department of Chemistry and Biochemistry,
UCLA, Los Angeles, CA 90095, USA}

\begin{abstract}

  We study the rheology and distribution of interparticle contact
  lifetimes for gravity-driven, dense granular flows of non-cohesive
  particles down an inclined plane using large-scale, three
  dimensional, granular dynamics simulations. Rather than observing a
  large number of long-lived contacts as might be expected for dense
  flows, brief binary collisions predominate. In the hard particle
  limit, the rheology conforms to Bagnold scaling, where the shear
  stress is quadratic in the strain rate. As the particles are made
  softer, however, we find significant deviations from Bagnold
  rheology; the material flows more like a viscous fluid. We attribute
  this change in the collective rheology of the material to subtle
  changes in the contact lifetime distribution involving the
  increasing lifetime and number of the long-lived contacts in the
  softer particle systems.
\end{abstract}

\maketitle

The rheology of granular materials is relevant to many areas of nature
and industry, from mountain avalanches and mud-slides, to grain
transport and storage \cite{savage1,iverson1}. We study a particularly
simple type of flow reminiscent of avalanches: gravity-driven, dense
granular flow down a rough, inclined plane.  This geometry is the
archetypal granular flow with which one can study the relation between
the stress state and the dynamics and structure, {\em i.e.} the
constitutive relation of the granular material.  Indeed, a number of
recent, well-controlled experiments
\cite{pouliquen1,ancey2,azanza1,ecke1}, and large-scale simulations
\cite{deniz2,leo7,leo19,taberlet1} have motivated numerous theories
that capture some of the features of inclined plane flows
\cite{mills4,ancey1,aranson1,bocquet1,lemaitre1,deniz3,louge2,kumaran1,jenkins1}.

To date, however, the most generally accepted treatment of granular
rheology is still the physical picture put forward by Bagnold over 50
years ago \cite{bagnold1,lois1}. Bagnold described a mechanism of
momentum transfer between particles in adjacent layers that assumes
instantaneous binary collisions between the particles during the flow.
Under this assumption, the inverse strain rate is the only relevant
time scale in the problem leading to a constitutive relationship
between the shear stress $\sigma$ and strain rate $\dot{\gamma}$ of
the form
\begin{equation}
\sigma = \kappa \dot{\gamma}^{2},
\label{eq1}
\end{equation}
where $\kappa$ is independent of ${\dot\gamma}$.

Despite recent concerns regarding the validity of Bagnold's original
experiments \cite{hunt1}, and the applicability of the theory
\cite{rajchenbach4}, Eq.~\ref{eq1} has proved rather successful in
predicting the rheology of dense granular flow down an inclined plane
\cite{pouliquen1,leo7}. This is somewhat at odds with the original
understanding of the Bagnold theory as it is based on the physics of
rapidly flowing hard-spheres where binary collisions predominate.
Dense flows \cite{vfacnote}, on the other hand, are thought to be
controlled by enduring, multiple interparticle contacts forming
extended stress-bearing structures \cite{mills4} and/or large
length-scale cooperative dynamics \cite{deniz3,pouliquen7,deniz4}.  In
fact, the stresses arising from such contact forces in dense flows are
practically an order of magnitude larger than the kinetic stresses
associated with velocity fluctuations \cite{leo7}.

In this letter, we analyse both the constitutive relation and the
interparticle contact dynamics of dense flows of non-cohesive granular
particles. To discuss interparticle dynamics we introduce two
timescales. The first is the lifetime of interparticle contacts and
the second is the inverse strain rate in the material. The inverse
strain rate or shear time sets the fundamental timescale over which
particle rearrangement events occur during the flow.

We find that the typical contact lifetimes in the system are of the
same order of magnitude as the binary collision time and insensitive
to the location of the pair of contacting particles within the flowing
pile. In contrast, the shear time is strongly dependent on the height
and varies considerably over the parameter space of interparticle
interactions studied here.  The scenario that emerges is somewhat
counter-intuitive: Even in dense flows the dynamics at the
microstructural scale remain predominately that of short-time, binary
collisions. When the interparticle contact lifetime is significantly
shorter than the shear time, the system is effectively in the
hard-particle limit, and the Bagnold constitutive relation, Eq. 1,
holds. Remarkably, the Bagnold relation is extremely sensitive to the
development of a small population of long-lived contacts on the
timescale of the shear time. Changes in the interparticle interaction
that lead to the enhancement of these longer-lived contacts also lead
to the breakdown of the Bagnold relation. In such cases it must be
supplemented by an additional term linear in the strain rate. Thus the
functional form of the constitutive law is sensitively dependent on
changes in the long-time tail of the interparticle contact time
distribution.

Our simulations are based on the model developed by Cundall and Strack
\cite{cundall1}, and Walton \cite{walton2}, and has been shown to
quantitatively match experimental data \cite{leo7,leo19,pouliquen1}.
We study $N=35,900$ monodisperse spheres of diameter $d$ and mass $m$,
flowing on a rough base of length $20d$, width $20d$, and tilted an
angle $\theta$ with respect to gravity. We use periodic boundary
conditions in the flow and vorticity directions. The height $h$, of
the flowing pile is between $90d < h < 100d$ depending on the angle of
inclination. For most of the results presented here, the inelastic
collisions are modelled as a Hookean spring and dashpot for forces both
normal ($n$) and tangential ($t$) to the interparticle contact plane,
but similar results were found for Hertzian contacts \cite{brewster2}.
At existing contacts we include a friction model that obeys the
Coulomb yield criterion with a coefficient of friction $\mu = 0.5$.

For Hookean contacts the coefficient of restitution parameterises the
dissipative nature of the interparticle collisions and is given by $
e_{\rm n} = \exp{(-\gamma_{\rm n} t_{\rm col}/2)}$. Here
$t_{\rm col}$ is the binary collision time,
\begin{equation}
t_{\rm col} =\pi [2k_{\rm n}/m - \gamma_{\rm n}^2/4]^{-1/2},
\label{eq2}
\end{equation}
where $k_{\rm n}$ and $\gamma_{\rm n}$ parameterise the elastic and
dissipative normal interparticle forces respectively. By
simultaneously adjusting these parameters, we fix $e_{\rm n}=0.88$ for
the results presented here. Varying $e_{\rm n}$ has little effect on
the lifetime distributions and the conclusions of this work. For the
tangential interactions we set $k_{\rm t} = 2 k_{\rm n}/7$ and
$\gamma_{\rm t}=0$.

We explore a range of spring constants $k_{\rm n}$, from
$10^{3}$--$10^{9}mg/d$. For $k_{\rm n} \leq 2 \times 10^{5}mg/d$,
the time step for numerical integrations $\delta t \sim 10^{-4} \tau$,
where time is in units of $\tau =\sqrt{d/g}$. $\delta t$ is adjusted
accordingly for larger $k_{\rm n}$, to ensure accurate dynamics. After
reaching a dynamical steady state, characterised by time-independent
total energy, where the mean flow velocity ranged from $10$--$100
\sqrt{dg}$ depending on the system parameters, we collected data over
a period of $\sim 10^{6} \delta t$.

\begin{figure}[h]
  \includegraphics[width=7cm]{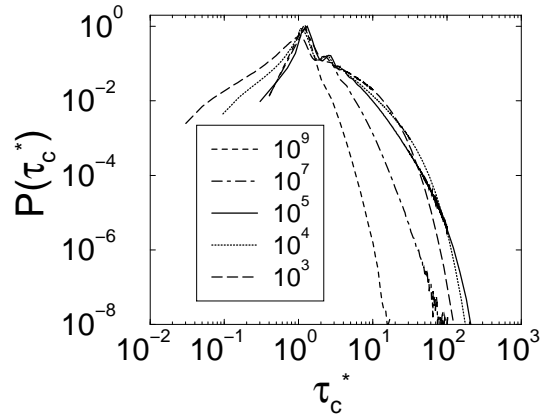}
  \caption  {Dependence of the distributions $P(\tau_{\rm{c}}^{*})$ of the
    depth-averaged normalised contact lifetimes, $\tau^{*}_{c} \equiv
    \tau_{\rm{c}} / t_{\rm col}$, on particle stiffness $k_{\rm{n}} =
    2 \times [ 10^{9}, ~ 10^{7}, ~ 10^{5}, ~ 10^{4}, ~ 10^{3} ] mg/d$.
    Data for $\theta = 23^{\circ}$, but the distributions are only
    weakly sensitive to $\theta$.}
    \label{fig1}
\end{figure}

In Fig.~\ref{fig1}(a) we show the distributions, $P(\tau^{*}_{c}
\equiv \tau_{c}/t_{\rm col})$, of two-particle contact lifetimes,
$\tau_{c}$, normalised by the binary collision time $t_{\rm col}$, for
different $k_{\rm n}$. Under this time rescaling all distributions
exhibit a prominent short-time peak near the binary collision time
$\tau_{c}^{*} = 1$, and an approximately exponential decay towards
longer contact lifetimes. As $k_{\rm{n}}$ is reduced this exponential
tail becomes broader indicating an increasing density of enduring
contacts. The data in Fig.~\ref{fig1} is depth-averaged over contacts
in the flowing pile away from the bottom boundary and the top
saltating layer, at an inclination angle, $\theta = 23^{\circ}$. The
distributions are qualitatively unchanged over the range of accessible
angles. We return to the more subtle dependence of $\tau_{\rm c}$ on
$\theta$ in our discussion of Fig.~\ref{fig5}.

\begin{figure}[h]
  \includegraphics[width=4cm]{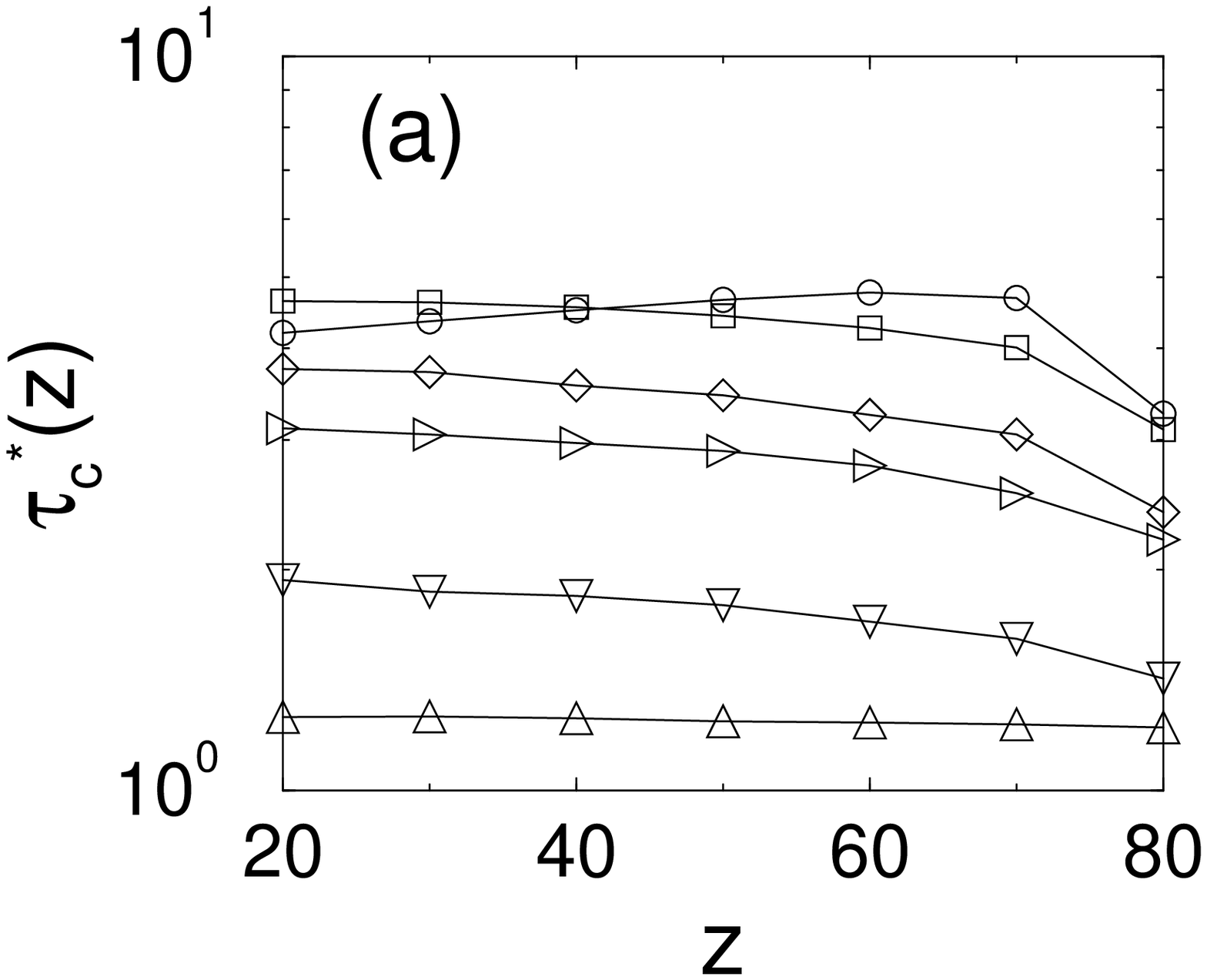}\hfill
  \includegraphics[width=4cm]{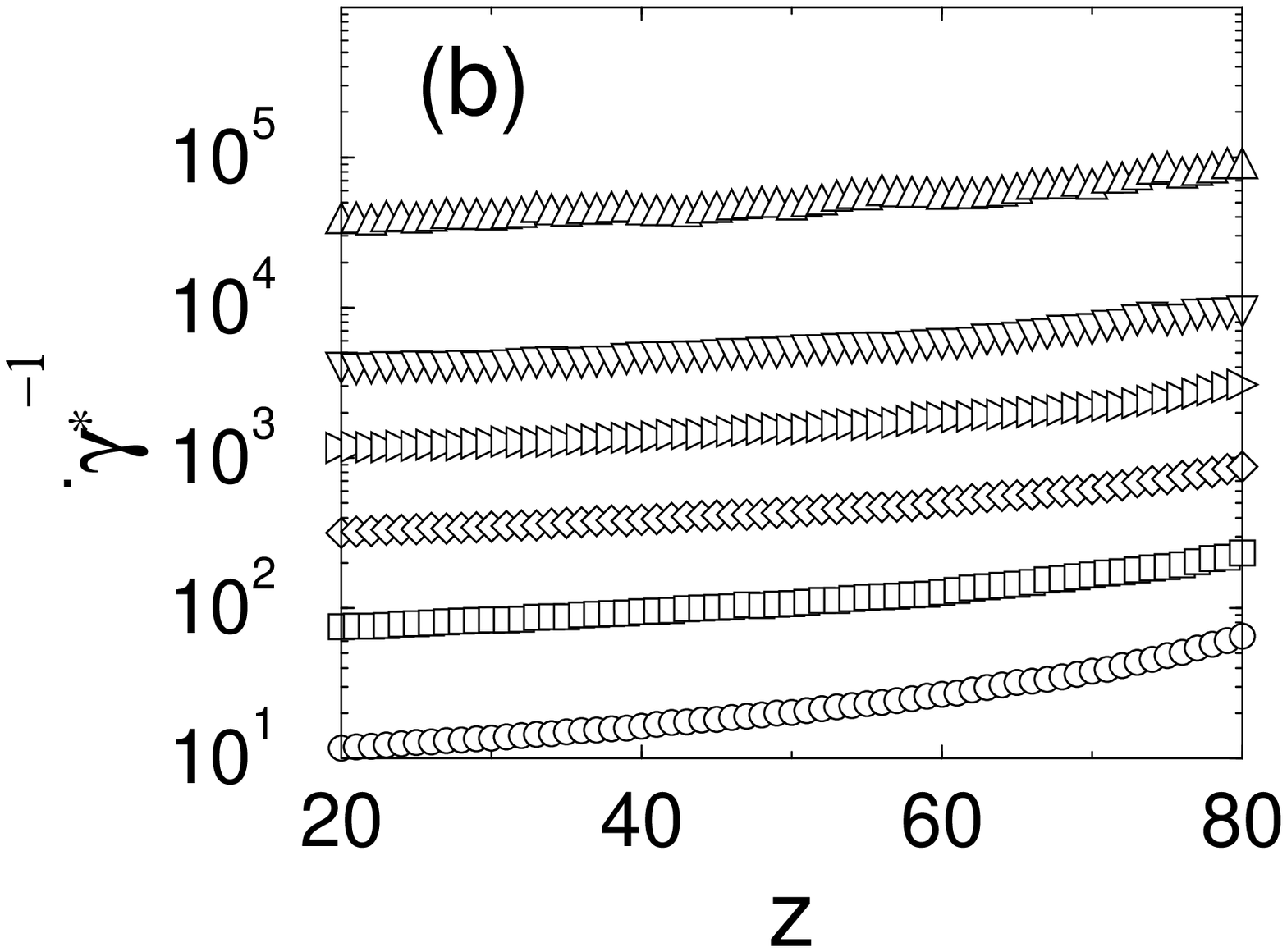}\hfill
  \caption{Depth profiles of, (a) the normalised contact time
    $\tau^{*}_{c}$ (line is a guide to the eye) and, (b) the
    normalised shear time $\dot{\gamma^{*}}^{-1}\equiv \dot\gamma^{-1}
    / t_{\rm col}$, for different particle stiffness, $k_{\rm{n}} = 2
    \times [ 10^{9} ~ (\triangle), ~ 10^{7} ~ (\triangledown), ~ 10^{6} ~
    (\rhd), ~ 10^{5} ~ (\Diamond), ~ 10^{4} ~ (\square), ~
    10^{3}(\circ) ] mg/d$. $z=0$ defines the bottom of the pile. Flow
    at $\theta=23^{\circ}$.}
    \label{fig2}
\end{figure}

The normalised mean contact lifetime exhibits essentially no depth
dependence as shown in Fig.~\ref{fig2}(a). Given the stress--free
boundary condition at the free surface of the pile, $\dot{\gamma}
\rightarrow 0$ there so the normalised inverse strain rate,
$\dot{\gamma^{*}}^{-1}\equiv \dot\gamma^{-1} / t_{\rm col}$, shown
in Fig.~\ref{fig2}(b), must be height dependent.  It also depends
strongly on $k_{n}$, varying by several orders of magnitude in our
data set.

\begin{figure}[ht]
  \includegraphics[width=4cm]{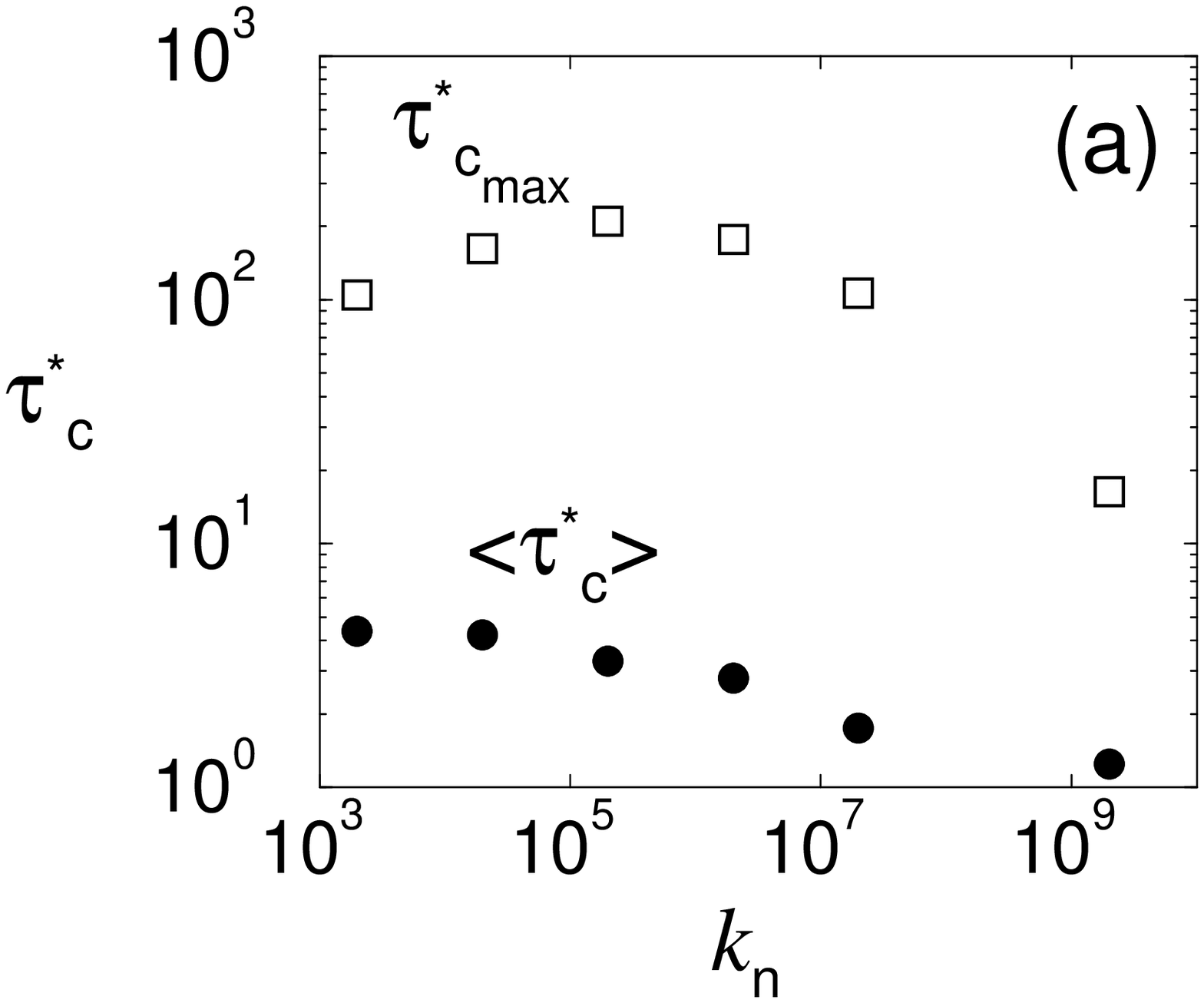}
  \includegraphics[width=4cm]{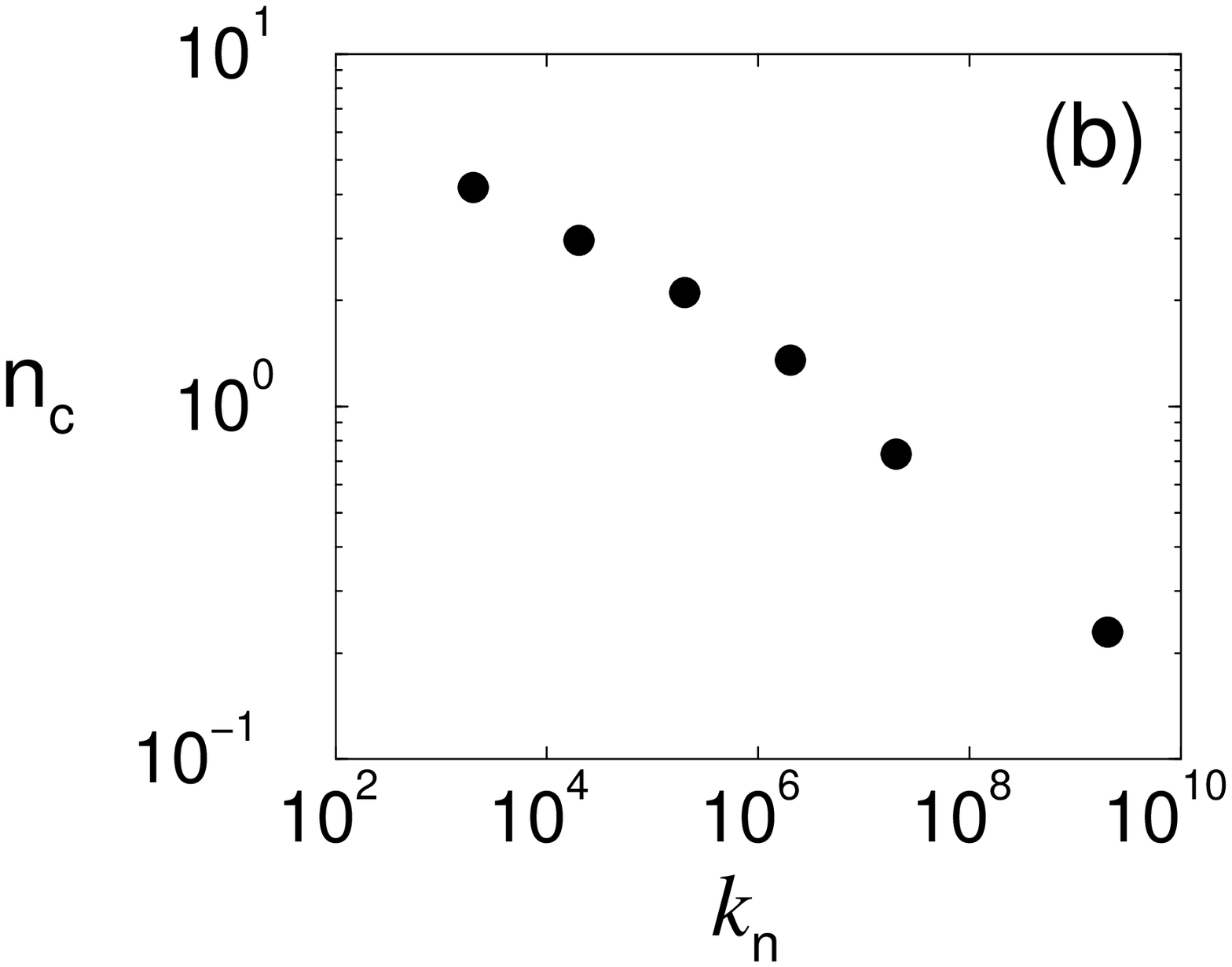}
  \caption{Dependence on the particle stiffness $k_{\rm n}$, of, (a) the
    depth-averaged contact lifetime $<\tau_{c}^{*}>$ (filled circles)
    and maximum contact time $\tau^{*}_{\rm{c_{max}}}$ (open squares),
    both normalised by $t_{\rm col}$, and (b) the depth-averaged number of
    contact neighbours per particle, $n_{c}$. Flow at
    $\theta=23^{\circ}$.}
  \label{fig3}
\end{figure}

As shown in Fig.~\ref{fig3}, over several orders of magnitude in
$k_{\rm{n}}$, the mean normalised contact time remains nearly constant
while the maximum contact time $\tau_{\rm{c_{max}}}$, extracted from
the distributions in Fig.~\ref{fig1}, decreases with increasing
stiffness, reflecting the narrowing of the contact time distributions
as the particles become harder. Additionally, the average number of
contacting neighbours per particle $n_{\rm{c}}$, shown in
Fig.~\ref{fig3}(b), tends to the binary collision limit ($ = 0$) as
the particles become harder. Thus, from a coordination point of view
the flows are not densely packed. Taken in combination, the data
presented in Figs.~\ref{fig1}--\ref{fig3} show that increasing the
stiffness of the interparticle interaction reduces the weight of the
exponentially small long-time tail of the contact lifetime
distribution. This narrowing of the contact lifetime distribution is
not observable from the more coarse measure of the mean contact
lifetime distribution, which remains close to the binary collision
time $t_{\rm col}$. Thus the particle dynamics are {\em always}
dominated by binary collisions, but changing the effective particle
stiffness causes subtle changes in the number of rare, enduring
contacts.

Since the interparticle contact lifetime is generically small compared
to the shear time, one might imagine that momentum transport is
dominated by effectively instantaneous binary collisions, and might
expect that the Bagnold constitutive law holds. To test this we now
turn to a characterisation of the granular rheology by fitting the
velocity profile of the flowing material to the prediction of a
modified Bagnold relation \cite{brewster1} of the form
\begin{equation}
  \sigma = \kappa \dot{\gamma}^{2} + \beta \dot{\gamma},
\label{eq3}
\end{equation}
where the coefficients $\kappa$ and $\beta$ are determined by a
least-squares fit to the velocity data. To characterise the deviations
from standard Bagnold rheology we define the ratio of the shear stress
in the Bagnold and linear forms to be $\Omega \equiv
\beta/\kappa\dot{\gamma}$.

\begin{figure}[h]
  \includegraphics[width=7cm]{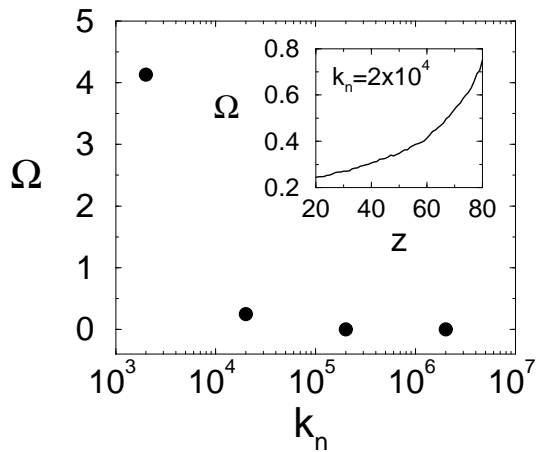}
  \caption{Dependence of the parameter $\Omega$ on the stiffness
    $k_{\rm n}$ for  $\theta=23^{\circ}$. Taken from the middle of
    the pile at $h/2$. Inset: Height dependence of $\Omega$, for
    $k_{\rm n} = 2\times 10^{4}mg/d$.}
  \label{fig4}
\end{figure}

\begin{figure}[h]
  \includegraphics[width=7cm]{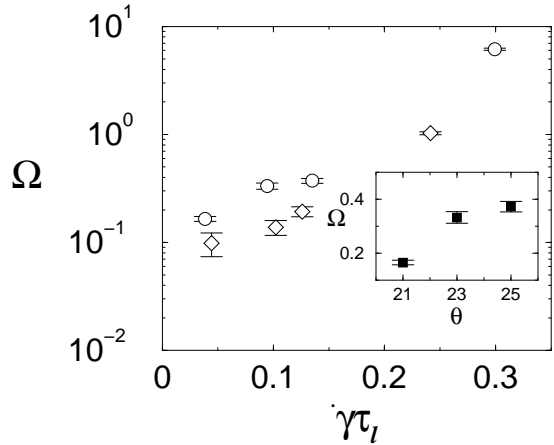}
  \caption{Plot of $\Omega$ vs. $\dot{\gamma} \tau_{l}$ as defined in
    the text for Hookean ($\circ$: $k_{\rm n} = \{10^{3},
    10^{4}\}mg/d$) and Hertzian ($\diamond$: $k_{\rm n} = \{10^4,
    10^5\}mg/d$) systems at inclination angles shown in the inset,
    which shows the dependence of $\Omega$ on $\theta$ on a linear
    scale.  The inclination angle is the most sensitive control of the
    shear rate $\dot{\gamma}$.}
  \label{fig5}
\end{figure}

Figure \ref{fig4} shows the dependence of $\Omega$ on $k_{\rm n}$ for
$\theta = 23^{\circ}$. In the hard-particle limit we expect $\Omega
\longrightarrow 0$. This is practically achieved already for
$k_{\rm{n}} \geq 2\times 10^{5} mg/d$, indicating that this value of
$k_{\rm{n}}$ is appropriate for modelling hard (glass) particles in a
system of this size \cite{leo7,leo19}. As the particles are made
softer, $\Omega$ grows meaning that the constitutive law approaches a
linear relation reminiscent of a viscous fluid. Since $\Omega$ is
inversely dependent on $\dot{\gamma}$, it is no surprise that it grows
monotonically as one approaches the free surface as is shown by the
inset to Fig.~\ref{fig4}. We have examined the (weaker) dependence of
$\Omega$ on all other particle interaction parameters, \emph{e.g.}
friction and inelasticity; these results will be discussed elsewhere
\cite{brewster2}.

For these dense flows the contact dynamics of \emph{all} systems
studied are dominated by binary collisions, that endure for no more
than $t_{\rm col}$, such that $\langle \tau_{\rm{c}} \rangle
\dot{\gamma} \ll 1$. But their rheology changes dramatically with
$k_{\rm n}$, going from the quadratic-in-shear-rate Bagnold law for
hard particles to a nearly linear or viscous relation for softer
particles. The dramatic change in the constitutive relation is
controlled by a more subtle feature of the contact time distribution
than simply the mean value. The significant rheological change appears
to be due to the growth of the lifetime and number of long-lived
contacts in the softer systems. We emphasise that the fraction of such
enduring contacts remains small in comparison to the more common
short-lived contacts with $\tau_{\rm{c}} \sim t_{\rm col}$.

The growth of a small population of long-lived contacts leads to the
formation of transient stress-bearing structures within the flowing
material, which can be rheologically significant. These structures
span streamlines in the flow and thus elastically transmit stress
across their length in proportion to the rate of particle impacts with
these structures $\sim \dot{\gamma}$. This picture is also consistent
with the rapid-quasi-static transition in shear flows
\cite{campbell6}. This reasoning suggests that the shear rate is the
appropriate local clock with which to measure contact lifetimes
$\tau_{\rm c}$ so that the dimensionless contact lifetime $\tau_{\rm
  c} \dot{\gamma}$ is the fundamental quantity controlling deviations
from Bagnold rheology in the flowing state. Given similar contact time
distributions, faster flows should then deviate more strongly from the
Bagnold prediction than slower flows.

To explore this point, we use the inclination angle of the pile to
adjust the overall shear rate and study the resulting change in the
observed rheology as parameterised by $\Omega$. In Fig.~\ref{fig5} we
plot $\Omega$ vs\ $\dot{\gamma} \tau_{l}$ for soft flows with
parameters spanning our range of stable angles and particle
stiffnesses. Here, $\tau_{l}$ is the lifetime of long-lived contacts
in the system, as defined by the criterion that these contacts are
sufficiently far out in the long-time tail of the normalised
contact-time distribution so that $P(\tau_{l}) = 10^{-3}$.  For these
systems, where we see pronounced deviations from the Bagnold
constitutive relation, $\Omega$ increases with increasing angle as
shown by the inset to Fig.~\ref{fig5}. All variations of $\Omega$ can
be collapsed onto a single master curve showing that the system's
rheology depends (exponentially) on the dimensionless contact lifetime
$\tau_l \dot{\gamma}$. The increasing value of these longer-lived
contacts measured relative to the fundamental clock-rate
$\dot{\gamma}$ drives the system from Bagnold to viscous rheology.
Simulations with $\Omega$ consistent with $0$ or $\infty$ are omitted
from Fig.\ref{fig5}, but follow the displayed trend.  Attempting to
similarly collapse all data sets using the dimensionless mean contact
time $\dot{\gamma} \langle \tau_c \rangle$ fails. Only the change in
the long-time tail of the contact distribution correlates with the
observed changes in the rheology.

In summary, we have provided extensive simulation results on the
rheology and interparticle contact statistics of gravity-driven,
dense, granular flows of non-cohesive grains. We observe a transition
from a Bagnold constitutive relation to one reminiscent of a Newtonian
fluid as the particles are made softer. Based on our numerical data,
and to account for this transition, we suggest a generalised Bagnold
relation \cite{brewster1} to quantify the changing rheology.
Furthermore, despite the na\"{i}ve guess that the flow in such dense
systems should predominantly involve long-lived, stress-bearing
structures, it turns out that for both hard and soft particles the
microscopic particle dynamics is dominated by frequent, short-time,
binary collisions. When examining the entire contact lifetime
distribution, however, we note that as the particles are made softer
there is a growth in the width of the distribution and that the
lifetime of rare, long-lived contacts grows relative to the shear
time. We propose that the emergence of these atypically long-lived
contacts is related to the dramatic change in the granular rheology
with particle stiffness and thereby rationalise the surprising result
that larger inclination angles, and hence faster flows are actually
less Bagnold in rheology than slower flows at smaller inclination
angles. It appears that the constitutive relation of granular media
interpolates between these extremes and is controlled by a combination
of the particle hardness and flow rate.

\acknowledgements
Sandia is a multiprogram laboratory operated by Sandia Corporation, a
Lockheed Martin Company, for the United States Department of Energy's
National Nuclear Security Administration under contract
DE-AC04-94AL85000.  RB and AJL gratefully acknowledge support for this
work from NASA 02-OBPR-03-C.

\end{document}